\documentclass[prb,english,reprint,superscriptaddress]{revtex4-2}
\usepackage[T1]{fontenc}
\usepackage[latin9]{inputenc}
\usepackage{bm}
\usepackage{amsmath}
\usepackage{amssymb}
\usepackage{bbold}
\usepackage{graphicx}
\usepackage{xcolor}

\definecolor{goodred}{rgb}{0.7,0,0}
\usepackage[colorlinks,urlcolor=blue,citecolor=blue,linkcolor=goodred]{hyperref}

\newcommand{\dd}[1]{\mathrm{d}#1\,}
\newcommand{\DD}[1]{\mathrm{D}[#1]\,}

\DeclareMathOperator{\Tr}{Tr}
\DeclareMathOperator{\tr}{tr}

\newcommand{\sgn}{\mathop{\mathrm{sgn}}}

\renewcommand{\vec}[1]{\bm{#1}}
\newcommand{\uvec}[1]{\hat{\bm{#1}}}

\graphicspath{{./figs/}}

\definecolor{PV-color}{rgb}{0.97,0.57,0.11}

\begin{document}

\title{Magnetoelectric effects in diffusive two-dimensional superconductors studied by the nonlinear $\sigma$ model}

\author{P. Virtanen}
\affiliation{Department of Physics and Nanoscience Center, University of Jyv\"askyl\"a, P.O. Box 35 (YFL), FI-40014 University of Jyv\"askyl\"a, Finland}
\email{pauli.t.virtanen@jyu.fi}

\begin{abstract}
  We describe a numerical approach to modeling magnetoelectric
  effects generated by spin-orbit coupling in inhomogeneous diffusive 2D
  superconductors. It is based on direct minimization of the free
  energy of diffusion modes, including their coupling to the
  spin-orbit field strength, described by a recently discovered
  $\sigma$-model action. We explain how to retain exact conservation
  laws in the the discretized model, and detail the numerical
  procedure. We apply the approach to the spin-galvanic and the
  inverse spin-galvanic effects in finite-size 2D superconductors, and
  describe short-range oscillations of circulating currents and spin
  densities originating from the spin-orbit coupling.
\end{abstract}

\maketitle

\section{Introduction}\label{sec:introduction}

Breaking simultaneously the inversion and the time-reversal symmetries
in a superconductor allows for several magnetoelectric effects.~\cite{bauer2012-ncs,smidman2017-sso} This includes spin-galvanic
coupling between charge and spin degrees of freedom,~\cite{edelstein1995-mep,yip2002-tds,edelstein2005-med,yip2014-ns,konschelle2015-tsg}
the $\varphi_0$-effect and helical phases,~\cite{buzdin2008-dcm,barzykin2002-isp,dimitrova2007-ttd,agterberg2007-mfi} and the supercurrent diode
effect~\cite{ando2020-osd,baumgartner2022-srm,wakatsuki2018-ncn,daido2022-isd,ilic2022-tsd,he2022-pts,davydova2022-ujd,nadeem2023-sde}.
One source for these effects arises from spin-orbit coupling (SOC)
enabled by inversion symmetry breaking in the material.
The theory for it in superconductors is most
developed for homogeneous bulk superconductors, and ballistic systems without disorder or systems where
Ginzburg--Landau type expansions are applicable.  The other limit of
disordered (``dirty'') inhomogeneous superconductors at low temperatures~\cite{eilenberger1968-tog,usadel1970-gde} has also been
extended to a description of different magnetoelectric effects originating
from SOC~\cite{houzet2015-qtd,konschelle2015-tsg,tokatly2017-uep,virtanen2022-nsm,kokkeler2022-ffa,ilic2024-sde}.

Recently in Refs.~\onlinecite{virtanen2021-mes,virtanen2022-nsm} we
suggested that various SOC-generated magnetoelectric effects in dirty
superconductors are captured by a $\sigma$-model~\cite{wegner1979-mep,efetov1980-idm,finkelshtein1983-ici,finkelshtein1987-stt,belitz1994-amt,feigelman2000-kad}
with one additional term in the action. However, despite the concise
nature of such formulations, they might appear a somewhat cumbersome
starting point compared to the quantum kinetic equations~\cite{eilenberger1968-tog} traditionally used in superconductivity
theory.

In this work we demonstrate that a formulation of the problem in terms of
the action instead of the kinetic equations arising from it is
immediately useful in numerical approaches, after a suitable
discretization, which is the only manual step.  The action formulation also
enables us to conveniently find a discretization that preserves local gauge
symmetries of the original model, and satisfies exact conservation
laws. We apply the method to the equilibrium spin-galvanic effect and
its inverse in 2D Rashba superconductors, and discuss resulting
spatial spin density oscillations at sample boundaries.

In Sec.~\ref{sec:model}, we outline the equilibrium version of the
theory of Ref.~\onlinecite{virtanen2022-nsm}, with technical details postponed
to Appendix~\ref{sec:matsubara}. In Sec.~\ref{sec:method} we formulate
a symmetry-preserving discretization and the numerical solution
strategy.  In Sec.~\ref{sec:magneto} we apply the numerical method to
spin-galvanic effects. Section~\ref{sec:summary} concludes the
discussion.

\section{Model}\label{sec:model}

We consider a metallic system, whose normal state Hamiltonian contains
a linear-in-momentum spin-orbit coupling (SOC) and an exchange field.
It can be written as
\begin{equation}
 \label{H-normalSO}
 H_0 = \frac{1}{2m}\Bigl(\vec{p} - \vec{A} - \frac{1}{2}\vec{\mathcal{A}}^{a}\sigma^a\Bigr)^2 - A_0 - \frac{1}{2}\mathcal{A}_0^a\sigma^a + V_{\rm imp}  \;. 
 \end{equation}
Here $\vec{p}$ is the momentum operator, $m$ the electron (effective)
mass, $\sigma^a$ are the Pauli matrices in spin space ($a=x,y,z$),
and $V_{\rm imp}$ is a scalar impurity potential. Summation over
repeated indices is implied. The electromagnetic vector potential is
$\vec{A}=(A_x,A_y,A_z)$ and the scalar potential is $A_0$. Also, $\vec{\mathcal{A}}^{a}=(\mathcal{A}_x^a,\mathcal{A}_y^a,\mathcal{A}_z^a)$ are the
SU(2) potentials~\cite{mineev1992-edm,mathur1992-qte,froehlich1993-gic,jin2006-sut,bernevig2006-ess,tokatly2008-esc,gorini2010-nag}
describing the spin-orbit coupling, and ${\cal A}_0^a$ describes the
exchange field in direction~$a$. We use here and below natural units with $e=\hbar=k_B=1$.

To describe superconductivity, the corresponding Bogoliubov--de Gennes
Hamiltonian is
\begin{align}
    \label{eq:Hstart}
    \mathcal{H}
    &=
    \tau_3
    \Bigl[
    \frac{(\vec{p} - \check{\vec{A}})^2}{2m} - \mu
    +
    V_{\rm imp}
    -
    \check{A}_0
    \Bigr]
    -
    \hat{\Delta}
    \\
    &=
    \mathcal{H}_0 + \tau_3 V_{\rm imp}
    \,,
\end{align}
where $\hat{\Delta} = \tau_+ \Delta + \tau_- \Delta^*$ and $\Delta$ is
the superconducting (singlet) pair potential, and we define $\check{A}_i = A_i
\tau_3 + \frac{1}{2}\mathcal{A}^a_i \sigma^a$ for $i=x,y,z$,
$\check{A}_0 = A_0 + \frac{1}{2}\mathcal{A}^a_0 \sigma^a\tau_3$.
Pauli matrices in the Nambu space are denoted $\tau_{1,2,3}$ and
$\tau_\pm=(\tau_1 \pm i\tau_2)/2$.  The above form corresponds to the Nambu--spin basis choice
$(\psi_{\uparrow},\psi_{\downarrow},\psi^\dagger_{\downarrow},-\psi^\dagger_{\uparrow})$.

We assume the random impurity potential $V_{\rm{}imp}$ is sufficiently
strong, so that the system is in the diffusive transport regime, where
the mean free path $\ell$ is still much longer than Fermi wavelength
$k_F^{-1}$, but much smaller than other length scales.

Finding the transport diffusion equation turns out to be convenient to
handle in the $\sigma$-model
formulation,~\cite{wegner1979-mep,efetov1980-idm,finkelshtein1983-ici,finkelshtein1987-stt,belitz1994-amt,feigelman2000-kad}
which has been used to study also various other features of the
disordered electron system. In this approach, the diffusion modes
of electrons are represented by a matrix field $Q$, and an action whose
saddle point equation can be interpreted
as the diffusion equation. With the spin-orbit fields included, in
addition to the diffusion terms, the action also contains a part
proportional to the electron-hole asymmetry of the dispersion. It
generates spin-Hall and other magnetoelectric
effects~\cite{virtanen2022-nsm} and the Hall
effect~\cite{levine1983-edm}. For equilibrium problems
in a semiclassical approximation, this description can be further
simplified to a form described below. Technical details can be found
in Appendix~\ref{sec:matsubara}.

The free energy for the model in Eq.~\eqref{eq:Hstart} averaged over
$V_{\rm imp}$ at temperature $T$ can be approximated based on the functional
\begin{align}
  \label{eq:action}
  F_0[Q]
  &=
  \sum_{\omega_n}
  \int\dd{r}
  \frac{\pi T}{8}
  \tr[\frac{\sigma_{xx}}{2}(\hat{\nabla}Q)^2 + 4i\nu_F\Omega Q
  \\\notag&\qquad
  - \frac{\sigma_{xy}'}{2} F_{ij}Q\hat{\nabla}_iQ\hat{\nabla}_jQ]
  \,.
\end{align}
Summation over repeated indices is implied.
Here, $\sigma_{xx}=2\nu_F{}D$ is the longitudinal Drude conductivity
where $D=\frac{1}{d}v_F\ell$ is the diffusion constant in $d$ dimensions;
$\nu_F$ is the density of states per spin projection at Fermi level and
$v_F$ the Fermi velocity. Also,
$\sigma_{xy}'=\frac{d\sigma_{xy}}{dB}\rvert_{B=0}=2\nu_F{}D\ell^2/(k_F\ell)$
is the zero-field derivative of the Hall conductivity.  The spin-orbit
coupling and vector potentials enter via the covariant derivatives $\hat{\nabla}_i =
\partial_{r_i} - i[\check{A}_i, \cdot]$ and the field strength
$F_{ij}=\partial_{r_i}\check{A}_j-\partial_{r_j}\check{A}_i-i[\check{A}_i,\check{A}_j]$.
Moreover, $\Omega = i\omega_n\tau_3 + \tau_3\hat{\Delta} +
\check{A}_0$, and $\omega_n=2\pi{}T(n+\frac{1}{2})$ are the Matsubara
frequencies.

The auxiliary field $Q(\vec{r},\omega_n)$ is a $4\times4$ matrix in
the Nambu--spin space, and it satisfies the condition
$Q(\vec{r},\omega_n)^2=1$ and the symmetry relations
\begin{subequations}\label{eq:Qsymmetry}
  \begin{align}
    Q(\vec{r},\omega_n) &= -\tau_3 Q(\vec{r},-\omega_n)^\dagger \tau_3
    \\
    &= \tau_1\sigma_yQ(\vec{r},-\omega_n)^T \sigma_y\tau_1
    \\
    \label{eq:Qlocalsym}
    &= -\tau_2\sigma_yQ(\vec{r},\omega_n)^* \sigma_y\tau_2
    \,.
  \end{align}
\end{subequations}
These properties are also satisfied by the momentum-averaged quasiclassical Green
function $g$~\cite{eilenberger1968-tog,usadel1970-gde,serene1983-qat,belzig1999-qgf}.

The value of the free energy is found by evaluating
Eq.~\eqref{eq:action} at the saddle point, $F=F_0[Q']$, where
$\delta{}F_0/\delta{}Q\rvert_{Q^2=1}=0$ at $Q=Q'$.  This saddle point
equation is the extension of the quasiclassical Usadel diffusion
equation~\cite{usadel1970-gde} to problems where weak spin-orbit coupling
and magnetoelectric effects are included~\cite{virtanen2022-nsm}.

\subsection{Other terms}

When the pair potential $\Delta$ in Eq.~\eqref{eq:Hstart} originates
from intrinsic superconductivity, the total free energy also contains
a superconducting mean-field term,
\begin{align}
  F_{\Delta} = \int\dd{^dr} \frac{|\Delta_{\vec{r}}|^2}{\lambda}
  \,,
\end{align}
where $\lambda$ is the interaction constant. The saddle-point equation
$\delta{}F/\delta\Delta=0$ with $F=F_0+F_\Delta$ produces the
self-consistency equation for the order parameter $\Delta$.  Here, we
only work within the BCS description of singlet superconductivity, but
this can be extended to more complex models.

The free energy can in principle also contain other terms allowed
by the symmetries of the underlying system, for example spin
relaxation~\cite{efetov1980-idm,abrikosov1960}
\begin{align}
  \label{eq:Fsiso}
  F_s[Q] = \frac{\pi\nu_FT}{8}\Tr[\frac{1}{\tau_s}(\tau_3\vec{\sigma} Q)^2 + \frac{1}{\tau_{\rm so}}(\vec{\sigma} Q)^2]
\end{align}
describing magnetic impurities and spin-orbit scattering, with
scattering times $\tau_s$ and $\tau_{\rm so}$.  Here we denoted
$\Tr=\sum_{\omega_n}\int\dd{r}\tr$.

Different types of boundaries of the system may also contribute
surface terms $F_b$ in the free energy, e.g.\ corresponding to tunnel
interfaces between materials.~\cite{altland2000-ftm,kupriyanov1988-iob,virtanen2016-scf} In particular,
vacuum boundary conditions have $F_b=0$, and can be considered by
restricting the space integrals to a finite volume.
Clean interfaces to large bulk reservoirs may be be modeled with a simpler
rigid-boundary approximation,~\cite{likharev79} where the value of
$Q(\vec{r})$ in some region is taken as fixed.  

\subsection{Observables}\label{sec:observables}

From the total free energy $F$ one can find observables: the charge $J^c$
and spin $J^s$~\cite{tokatly2008-esc} currents are found by taking
derivatives with respect to the potentials,
\begin{align}
  J^c_i
  &=
  -\frac{\delta F}{\delta A_i}
  =
  \frac{i\pi T}{4}\sum_{\omega_n}\tr\tau_3\mathcal{J}_i
  \,,
  \\
  \label{eq:jspin}
  J^s_{ia}
  &=
  -\frac{\delta F}{\delta \mathcal{A}_i^a}
  =
  \frac{i\pi T}{8}\sum_{\omega_n}\tr\sigma^a\mathcal{J}_i
  \,,
  \\
  \label{eq:Jdef}
  [\mathcal{J}_\mu(\vec{r},\omega_n)]_{\alpha\beta}
  &\equiv
  -
  \frac{4}{i\pi T}
  \frac{\delta F(\omega_n)}{\delta [\check{A}_\mu(\vec{r})]_{\beta\alpha}}
  \,,
\end{align}
where $\mu\in\{0,x,y,z\}$, $i,j\in\{x,y,z\}$, and $\alpha,\beta$ are
matrix indices in the Nambu--spin basis. Here, $\mathcal{J}$ is
defined as a derivative of a single term of the Matsubara sum in
Eq.~\eqref{eq:action}, in such a way that it is equivalent with the ``matrix
current''~\cite{nazarov94} in quasiclassical theory, e.g.\ if omitting the
spin-orbit terms,
$\mathcal{J}_i=-\sigma_{xx}Q\partial_{r_i}Q$, $\mathcal{J}_0=-2\nu_FQ$.

Similarly, the local spin and
charge accumulations are given by
\begin{align}
  \label{eq:Sdef}
  S^a
  =
  -\frac{\delta F}{\delta \mathcal{A}_0^a}
  =
  \frac{i\pi T}{8}\sum_{\omega_n}\tr\sigma^a\tau_3\mathcal{J}_0
  =
  \frac{\pi \nu_F T}{4i}\sum_{\omega_n}\tr\sigma^a\tau_3Q
  \,,
  \\
  \delta\rho
  =
  -\frac{\delta F}{\delta A_0}
  =
  \frac{i\pi T}{4}\sum_{\omega_n}\tr{}\mathcal{J}_0
  =
  \frac{\pi \nu_F T}{2i}\sum_{\omega_n}\tr{}Q
  =
  0
  \,.
\end{align}
In general $\delta\rho=0$ in the equilibrium situation, as we are
considering the metallic regime where the system is charge neutral.

Note that $F_0$ and the above expressions for the observables relate
to the low-energy diffusion modes. Similarly as in the quasiclassical
theory, there is a second contribution involving the full
electron band. In this model, it does not directly couple to the low-energy physics
but produces the normal-state equilibrium currents and densities
(see~Appendix~\ref{sec:matsubara}).

The free energy $F_0$ is invariant in transformations that change the
choices of the electromagnetic gauge and the spin quantization
axis, which implies conservation laws.~\cite{tokatly2008-esc,jin2006-sut,mineev1992-edm,mathur1992-qte} Here,
this means the transformations
\begin{subequations}\label{eq:gaugetrans}
\begin{align}
  Q({\vec{r}})&\mapsto{}W_{\vec{r}}Q({\vec{r}})W_{\vec{r}}^\dagger
  \,,
  \quad
  W_{\vec{r}} = e^{i\phi_{\vec{r}}\tau_3 + i\theta^a_{\vec{r}}\sigma^a}
  \,,
  \\
  \check{A}_{j\vec{r}}&\mapsto{}W_{\vec{r}}\check{A}_{j\vec{r}}W_{\vec{r}}^\dagger-i(\partial_{r_j}{}W_{\vec{r}})W_{\vec{r}}^\dagger
  \,,
  \quad
  j=x,y,z
  \,,
\end{align}
\end{subequations}
and
$\check{A}_{0\vec{r}}\mapsto{}W_{\vec{r}}\check{A}_{0\vec{r}}W_{\vec{r}}^\dagger$,
where the angles $\phi$, $\theta$ are arbitrary. The invariance
results to~\cite{tokatly2008-esc}
\begin{subequations}\label{eq:conservation-laws}
\begin{align}
  0
  &=
  \partial_{r_i} J^c_i
  \,,
  \\
  \label{eq:Jsconserv}
  0
  &=
  \partial_{r_i} J_{ia}^s + \epsilon_{abc}\mathcal{A}_i^b J^s_{ic} + \epsilon_{abc}\mathcal{A}_0^b S^c
  \,,
\end{align}
\end{subequations}
describing charge and covariant spin conservation.  Here,
$\epsilon_{abc}$ is the Levi--Civita symbol.

\section{Method}\label{sec:method}

Given the free energy functional $F[Q]$, one can then access
equilibrium magnetoelectric effects in diffusive systems.
We now wish to find the values $Q$ that solve
the saddle-point equation:
\begin{align}
  \frac{\delta F}{\delta Q}
  \biggr\rvert_{Q^2=1}
  = 0
  \,.
\end{align}
In addition, we need to compute the values of the currents and
densities by evaluating the corresponding derivatives. We solve the
problem numerically (implementation is available~\cite{code}), which
requires discretization of the continuum action.

\subsection{Gauge-invariant discretization}

Naive discretization does not preserve the
conservation law of the spin current. Moreover, given the presence of
the magnetoelectric coupling (Hall term), also charge conservation can
be broken.  Although the magnitude of the discretization artifacts in
the conservation laws should in general decrease as the lattice
spacing is reduced, this can occur slowly.  To preserve conservation
laws exactly, it is advantageous to formulate the problem in such a
way that a discrete version of the gauge invariance is retained.  This
is essentially similar to using the Peierls substitution in a
tight-binding model, and a related approach is commonly used in lattice
gauge theories~\cite{gupta1998-ilq}.

We discretize the action as follows. We subdivide the space to
rectangular cells, centered on a lattice $\vec{r}_j=(h j_x, h j_y, h
j_z)$ with $j_{x,y,z}\in\mathbb{Z}$ and $h$ is the lattice spacing.
We choose $Q_j = Q(\vec{r}_j)$ to be the values of $Q$ at the lattice
sites. Similarly,
$\check{A}_{0j}=\check{A}_0(\vec{r}_j)$, $\hat{\Delta}_j=\hat{\Delta}(\vec{r}_j)$,
and $\Omega_j=i\omega\tau_3+\check{A}_{0j} + \tau_3\hat{\Delta}_j$.

The translation associated with the gauge invariant derivative,
$e^{(\vec{r}_j-\vec{r}_i)\cdot\hat{\nabla}}Q(\vec{r}_i) =U_{ij}Q(\vec{r}_j)U_{ji}$,
contains phase factors in addition to the spatial translation.~\cite{elze1989-qgt}
These are the Wilson link matrices
\begin{align}
  U_{ij} =U_{ji}^{-1} =
  \mathrm{Pexp}\Bigl[i\int_{L(\vec{r}_i,\vec{r}_j)}d\vec{r}'\,\cdot\check{\vec{A}}(\vec{r}')\Bigr]
  \,,
\end{align}
where $L(\vec{r}_i,\vec{r}_j)$ is the straight line from $\vec{r}_j$
to $\vec{r}_i$ and $\mathrm{Pexp}$ is the path-ordered integral.
Under gauge transformation~\eqref{eq:gaugetrans},
$U_{ij}\mapsto{}W_{\vec{r}_i}U_{ij}W_{\vec{r}_j}^{\dagger}$.  The
expression
$\frac{1}{h}[e^{(\vec{r}_j-\vec{r}_i)\cdot\hat{\nabla}}-1]Q(\vec{r}_i)=\frac{1}{h}[U_{ij}Q_jU_{ji}-Q_i]$
then has a local transformation
$(\ldots)\mapsto{}W_{\vec{r}_i}(\ldots)W_{\vec{r}_i}^{\dagger}$, and
its expansion for $h=|\vec{r}_i-\vec{r}_j|\to0$ produces the
gauge-invariant derivative in direction $\vec{r}_j-\vec{r}_i$ at point
$\vec{r}_i$.  The field strength $F_{\mu\nu}$ can also be produced
similarly from combinations of $U$.~\cite{gupta1998-ilq} All
derivative parts of the action can be constructed from $U$ and on-site
$Q$, and the result can be made gauge invariant by ensuring spatially
separated points are connected by $U_{ij}$.

Hence, for the discretization of the free energy $F_0$, we require
invariance under the discrete transformations
\begin{align}
  \label{eq:discrete-gauge-invariance}
  Q_j &\mapsto u(j) Q_j u(j)^{-1} \,,
  &
  U_{ij}
  &\mapsto
  u(i)U_{ij} u(j)^{-1}
  \,,
  \\\notag
  \Omega_j &\mapsto{}u(j)\Omega_ju(j)^{-1}
  \,,
\end{align}
for any set of transformation matrices $u(i)$ on each site.

We can then start constructing a discretization satisfying
Eq.~\eqref{eq:discrete-gauge-invariance}. We discretize the directional gradient
$\vec{e}_{ij}\cdot\hat{\nabla}Q$ along $\vec{e}_{ij}=(\vec{r}_i-\vec{r}_j)/|\vec{r}_i-\vec{r}_j|$
on a link between cells $i$ and $j$ as
\begin{align}
  D_{ij}
  :=
  \frac{1}{h}(Q_i U_{ij} - U_{ij} Q_j)
  \,.
\end{align}
This choice is useful, because in addition to the simple
transformation law $D_{ij}\mapsto{}u(i)D_{ij}u(j)^{-1}$, it retains an
exact ``anticommutation'' relation $Q_i D_{ij} = - D_{ij} Q_j$,
analogous to the property $Q\hat{\nabla}Q=-(\hat{\nabla}Q)Q$ of the
continuum derivative that arises due to the normalization
$Q^2=1$. This ensures that the various equivalent continuum forms that can be
constructed by reordering $Q$ and $\hat{\nabla}Q$ are also equivalent
in the discretized case.  For $h\to0$, one can verify that
the above produces the correct continuum limit: $D_{ij}\to\frac{Q_i -
  Q_j}{h} -
[i\vec{e}_{ij}\cdot\check{\vec{A}}(\frac{\vec{r}_i+\vec{r}_j}{2}),
  Q_j]\to\vec{e}_{ij}\cdot\hat{\nabla}Q$.

We then discretize
\begin{align}
 \int\dd{r}\tr (\hat{\nabla} Q)^2
 \mapsto
 -h^d \sum_{\mathrm{neigh}(i,j)} \tr D_{ij} D_{ji}
\end{align}
where $d$ is the space dimension and $\mathrm{neigh}(i,j)$ indicates summation over
links between neighbors on the lattice.  From the transformation
properties of $D_{ij}$, the above is invariant under
Eq.~\eqref{eq:discrete-gauge-invariance}.

\begin{figure}[t]
  \includegraphics{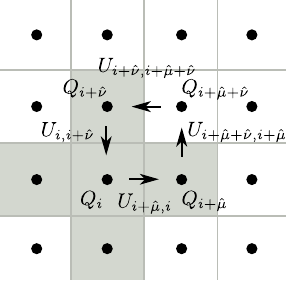}
  \caption{\label{fig:discretization}
    Lattice discretization.
    Neighbors of site $i$ are shaded,
    and the arrows indicate the plaquette loop $P_{lkji}$
    for the coordinate directions $\hat{\mu}$, $\hat{\nu}$.
  }
\end{figure}

The discretization of the field-strength $F_{\mu\nu}$ term can be made following
same ideas as in lattice QCD, where the gluon action is expressed
in terms of the link matrices $U$ via a plaquette loop~\cite{gupta1998-ilq}.  Consider
the plaquette (see Fig.~\ref{fig:discretization}), with corner site $i$
and axes $\mu\ne\nu$, and denote $j=i+\hat{\mu}$, $k=i+\hat{\mu}+\hat{\nu}$, $l=i+\hat{\nu}$.
Expand around $\vec{r}_0=\frac{\vec{r}_i+\vec{r}_j+\vec{r}_k+\vec{r}_l}{4}$:
\begin{align}
   P_{lkji}
   &:=
   U_{il}U_{lk}U_{kj}U_{ji}
   =
   1
   + i h^2 F_{\mu \nu}(\vec{r}_0)
   +
   \mathcal{O}(h^3)
   \,,
\end{align}
which then provides an expression for $F$ in terms of the link matrices $U$.
The above expansion follows from the continuum limit expansion of the link matrices
\begin{align}
 U_{ji}
 &=
 \sum_{n=0}^\infty i^n |\vec{r}_i-\vec{r}_j|^n \int_{-1/2}^{1/2}ds_1\int_{-1/2}^{s_1}ds_2\ldots\int_{-1/2}^{s_{n-1}}ds_n\,
 \notag
 \\&\qquad\qquad\times\mathcal{A}(s_1)\cdots\mathcal{A}(s_n),
\end{align}
where $\mathcal{A}(s)=\vec{e}_{ij}\cdot\check{\vec{A}}[(\frac{1}{2} - s)\vec{r}_i +
  (\frac{1}{2} + s)\vec{r}_j]$.

We can then express
\begin{align}
  \tr
  F_{\mu\nu}Q \hat{\nabla}_\mu Q \hat{\nabla}_\nu Q
  &\mapsto
  \frac{i}{h^4}
  \tr
  (1 - P_{lkji}) U_{ij} D_{ji} Q_i D_{il} U_{li}
  \,.
\end{align}
This expression now both retains the discrete gauge invariance, and in
the continuum limit reduces to the correct term.  Summation over the
$\mu$, $\nu$ indices must also be done. To avoid directionality bias,
we express it as an average over the corners of the
plaquettes as follows. This gives a discretization of the Hall term
$F_H=\int\dd{r} \tr F_{ij} Q\hat{\nabla}_iQ\hat{\nabla}_jQ$:
\begin{align}
  F_H
  &\mapsto
  \frac{h^d}{2ih^4}
  \sum_{\mathrm{plaqc}(ijkl; i)}
  \tr
  (U_{lk} U_{kj} - U_{li} U_{ij}) D_{ji} Q_i D_{il}
  \,,
\end{align}
where $\mathrm{plaqc}(ijkl; i)$ implies summation over all counterclockwise
plaquettes surrounding all corner sites $i$.

The discretization of the remaining local terms is straightforward,
and can be done as
\begin{align}
  \int\dd{r}\tr \Omega Q\mapsto\sum_ih^d\tr(\Omega_i Q_i)
  \,.
\end{align}
One can discretize $F_s$ in the same way.

\subsection{Discrete conservation laws}

Consider then the discrete conservation laws that are present in the discretized
free-energy functional $F=F_0[Q,U,\Delta,\check{A}_0]+F_\Delta[\Delta]+F_s[Q]$, where $F_0$ has the
symmetries~\eqref{eq:discrete-gauge-invariance}.

From the definition of the current as a derivative with respect to the
gauge potentials~\eqref{eq:Jdef}, one finds the discrete current
incoming from cell $j$ measured at site $i$ along the link $(i,j)$:
\begin{gather}
  \label{eq:lattice-J-def}
  J_{ij}^a
  =
  \frac{\partial}{i\partial\xi}
  F_0[U^{(i,j)}, Q, \Delta, \check{A}_0]\rvert_{\xi=0}
  \,,
  \\
  \label{eq:lattice-JU-def}
  U^{(i,j)}:
  \quad
  U_{ij}
  \mapsto
  e^{i\xi T^a}
  U_{ij}
  \,,
  U_{ji}
  \mapsto
  U_{ji}
  e^{-i\xi T^a}
  \,,
\end{gather}
where $T^a$ is an appropriate matrix generator of the current,
$T^0=\tau_3$ for the charge current and $T^{x,y,z}=\sigma_{x,y,z}$ for the spin current,
and the
transformation is made only in the link $(i,j)$.

The local gauge invariance~\eqref{eq:discrete-gauge-invariance} then
implies that the model has a discrete continuity equation
\begin{gather}
  \sum_{j\in{}\mathrm{neigh}(i)} J_{ij}^a
  =
  R_i^a
  \equiv
  \frac{\partial}{i\partial\xi}
  F_0[U,Q^{(i)}, \Delta^{(i)},\check{A}_0^{(i)}]\rvert_{\xi=0}
  \,,
  \\
  Q^{(i)}:
  \quad Q_i\mapsto{}e^{-i\xi T^a}Q_ie^{i\xi T^a},
  \\
  \hat{\Delta}^{(i)}:
  \quad
  \hat{\Delta}_i\mapsto{}e^{-i\xi T^a}\hat{\Delta}_ie^{i\xi T^a}
  \,,
  \\
  \check{A}_0^{(i)}:
  \quad
  \check{A}_{0i}\mapsto{}e^{-i\xi T^a}\check{A}_{0i}e^{i\xi T^a}
  \,,
\end{gather}
which defines the current sink term $R_i^a$.

Given a functional for the total free energy, $F=F_0+F_\Delta+F_s$,
using the saddle point conditions $\delta F/\delta{}Q=0$ and $\delta F/\delta{}\Delta=0$
we can rewrite
\begin{align}
  R_i^a
  &=
  \frac{\partial}{i\partial\xi}
  F_0[U,Q,\Delta,\check{A}_0^{(i)}]\rvert_{\xi=0}
  \\\notag
  &
  -
  \frac{\partial}{i\partial\xi} F_s[Q^{(i)}]\rvert_{\xi=0}
  -
  \frac{\partial}{i\partial\xi} F_\Delta[\Delta^{(i)}]\rvert_{\xi=0}
  \,.
\end{align}
The last term vanishes due to invariance of $F_\Delta$.
For $a=0$, also the other terms vanish, so charge is conserved in the
discrete model, $R^0_i=0$.

For $a=x,y,z$ the remaining terms describe the divergence of spin
current: the exchange field term --- corresponding to the last term in
Eq.~\eqref{eq:Jsconserv} --- and the spin relaxation. Note that for
the isotropic spin relaxation of Eq.~\eqref{eq:Fsiso},
$F_s[Q^{(i)}]=F_s[Q]$, so that it is not a sink or source for
\emph{equilibrium} spin currents.

The second term in the covariant conservation law~\eqref{eq:Jsconserv}
also appears in the discrete formulation: The
definition of the lattice spin current is not symmetric,
$J_{ij}\ne{}-J_{ji}$. From Eq.~\eqref{eq:lattice-JU-def} one can
observe that the asymmetry is present only when $T^a$ does not commute
with $U_{ij}$. This is an issue only for the spin current, as the
charge current generator $T^0=\tau_3$ always commutes with
$U_{ij}$. What this means is that since the parallel transport of spin
between the sites $i$ and $j$ can imply spin rotation described by the
gauge field, corresponding to the second term in
Eq.~\eqref{eq:Jsconserv}, the currents measured at separated points
are generally not equal.

\subsection{Numerical implementation}

Manually evaluating the derivatives $\delta F/\delta{}Q$ to find the
saddle-point (Usadel) equations is somewhat unwieldy.  To avoid it,
we can make use of algorithmic differentiation methods. The advantage
here is that they require as an input only a computer routine
evaluating the discretized free energy functional $F$, given variables
$\{Q_j\}$ as input.  The first (gradient) and second (Hessian)
derivatives of the discretized $F$ with respect to the input variables
can then be automatically deduced, allowing the use of efficient
gradient-based optimization methods.

In practice, all manual symbolic manipulation necessary is then
already completed in the previous section. Hence, we have an
essentially \emph{action-based} numerical method, where it would be
very simple to e.g.\ include additional terms in Eq.~\eqref{eq:action}
if needed.  The discrete gauge invariance also implies the approach is
not sensitive to the gauge choices, and satisfies conservation laws
exactly.

We use the CppAD library~\cite{cppad} for computing the gradient and
Hessian of $F$ with respect to the input variables. Moreover, it is
also used to evaluate the action derivatives giving the currents~\eqref{eq:Jdef}.

The Matsubara sums are evaluated using a Gaussian sum quadrature~\cite{monien2010-gqs},
\begin{align}
  \sum_{n=-\infty}^\infty f(\omega_n)
  \simeq
  \sum_{j=1}^N \tilde{a}_j f(\tilde{\omega}_j(T))
  \,,
\end{align}
where $\tilde{a}_j$ and $\tilde{\omega}_j(T)=2\pi
T(\tilde{n}_j+\frac{1}{2})$ are such that the equality is exact
for any functions $f(\omega_n)$ that are polynomials in $(1+|n|)^{-2}$ of
order less than $N/2$. Here, $N$ is chosen such that the largest
$\tilde{\omega}_j$ is much larger than the typical energy scale of
$Q(\omega)$. The quadrature is generally more rapidly convergent
than naive summation, and moreover ensures that also large values of
$\omega$ are sampled without needing very large $N$.

The saddle-point equation $\delta{}F/\delta{}Q=0$ is solved with a
preconditioned Newton--Krylov method.~\cite{knoll2003-jnm}
Parametrizing $\{Q_i\}$ in terms of real variables $\vec{x}=\{x_i\}$,
this corresponds to iteration
$\vec{x}^{(n+1)}=\vec{x}^{(n)}+\delta\vec{x}^{(n)}$, where a Krylov
method solves the problems
\begin{align}
  M\frac{\partial^2 F[x^{(n)}]}{\partial x \partial x} \delta \vec{x}^{(n)}
  &=
  -M\frac{\partial F[x^{(n)}]}{\partial x}
  \,,
\end{align}
where $\partial F/(\partial x \partial x)$ is the Hessian and
$\partial{}F/\partial{}x$ the gradient of $F$ with respect to the real
input variables. We use a number of standard approaches to accelerate
the solution.  The preconditioner $M$ is taken to be the
(incomplete) sparse LU inverse~\cite{davis2004-a8u} of the Hessian.
Updating the Hessian is an  expensive
step in the calculation, and in general we keep $M$ ``frozen'' for
several iterations and recompute it only if the Krylov convergence
starts to suffer.  Computation of the Hessian is avoided in the Krylov
steps themselves, as they only require the matrix-vector products
$(\partial
F/\partial{}x\partial{}x)\vec{y}\simeq\frac{1}{\alpha}\bigl(\frac{\partial{}F}{\partial{}x}[\vec{x}^{(n)}+\alpha \vec{y}] -
\frac{\partial F}{\partial{}x}[\vec{x}^{(n)}]\bigr)$, which can be
computed as numerical derivatives of the gradient with $\alpha\to0$.

The conditions $Q_i^2=1$ are here eliminated by a Riccati
parametrization~\cite{schopohl1995-qsa} of the matrix $Q_i$, in terms
of unconstrained complex $2\times2$ matrices $\gamma$ and
$\tilde{\gamma}$:
\begin{align}
  \label{eq:riccati}
  Q_i =
  \sgn(\omega_n)
  \begin{pmatrix}
    N_i & 0
    \\
    0 & \tilde{N}_i
  \end{pmatrix}
  \begin{pmatrix}
    1 - \gamma_i\tilde{\gamma}_i & 2\gamma_i
    \\
    2\tilde{\gamma}_i & -1 + \tilde{\gamma}_i\gamma_i
  \end{pmatrix}
  \,,
\end{align}
where $N_i=(1 + \gamma_i\tilde{\gamma}_i)^{-1}$ and $\tilde{N}_i = (1 + \tilde{\gamma}_i\gamma_i)^{-1}$.
One can also use the symmetry~\eqref{eq:Qlocalsym} which implies
$\tilde{\gamma}_i = \sigma_y\gamma_i^*\sigma_y$ for real $\omega_n$ to reduce the number of
variables further. As this symmetry comes from a symmetry of the action,
it is also possible to leave $\tilde{\gamma}_i$ free in which case
the symmetry is implicitly included in the saddle point equations.
The real variables $\vec{x}$ are then the real
and imaginary parts of the matrix elements of $\gamma_i$ (and
$\tilde{\gamma}_i$ if left in) in each cell in the discretization.

\subsection{BCS cutoff}\label{sec:bcs-cutoff}

To deal with the BCS cutoff, we use the usual cutoff elimination by
adding and subtracting
\begin{align}
  &
  F_{\Delta}
  -\frac{\pi\nu_FT}{8}\Tr[4\hat{\tau}_3(\omega - i\hat{\Delta})Q]
  =
  \mathrm{const.}
  + F_{\Delta}'
  \\&\quad\notag
  -\frac{\pi\nu_FT}{2}\Tr\bigl[
    \hat{\tau}_3(\omega - i\hat{\Delta})(Q-\tau_3\sgn\omega)
    -
    \frac{|\Delta|^2}{2|\omega|}
  \bigr]
 \,,
\end{align}
where the part in $\Tr[\ldots]$ contains a convergent Matsubara sum at the saddle point $Q_*$
as the slowly decaying part in $Q_*$ at $|\omega_n|\to\infty$ is canceled. This results to
\begin{align}
  F_{\Delta}'
  &=
  \nu_F (\frac{1}{\nu_F\lambda} - \pi{}T\sum_{|\omega_n|<\omega_c}\frac{1}{|\omega_n|})\int\dd{^dr} |\Delta|^2
  \\
  &\simeq
  \nu_F \log\big(\frac{T}{T_{c0}}\bigr)
  \int\dd{^dr} |\Delta|^2
  \,,
\end{align}
where $T_{c0}=(2e^{\gamma}\omega_c/\pi)e^{-1/(\lambda\nu_F)}=(e^{\gamma}/\pi)\Delta_0$ is the BCS critical temperature and
$\gamma$ is the Euler constant.

\section{Magnetoelectric response}\label{sec:magneto}

\begin{figure}
  \includegraphics{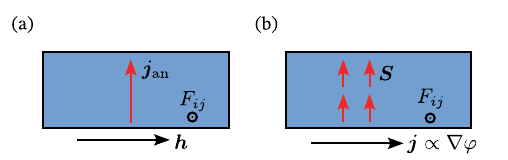}
  \caption{\label{fig:sge}
    (a) Spin-galvanic effect: anomalous charge current $\vec{j}_{\rm an}$
    generated by the Rashba SOC field strength $F_{ij}$ and an exchange field $\vec{h}$.
    (b) Inverse spin-galvanic effect: spin density $\vec{S}$ generated by
    $F_{ij}$ and charge supercurrent $\vec{j}$.
    Both effects are perturbed by finite sample boundaries.
  }
\end{figure}

The spin-orbit coupling results to several magnetoelectric effects in
superconductors. One of them is the spin-galvanic or inverse Edelstein effect, generation of
anomalous equilibrium supercurrents due to external Zeeman fields, and
its inverse effect.~\cite{edelstein1995-mep,edelstein2005-med,yip2002-tds,dimitrova2007-ttd}
These are schematically illustrated in Fig.~\ref{fig:sge}.

\subsection{Spin-galvanic effect}

Consider a finite-size 2D superconducting layer of size $L\times{}W$,
with Rashba spin-orbit coupling $\check{A}_{x}=\alpha\sigma_y$,
$\check{A}_y=-\alpha\sigma_x$, $F_{ij}=2\alpha^2\sigma_z\epsilon_{ijz}$,
and an internal exchange field
$\vec{h}=h\hat{x}$. In this configuration, the magnetoelectric
coupling generates equilibrium supercurrents, whose flow is restricted
by the sample boundaries.  These effects were previously theoretically
studied in Ref.~\onlinecite{bergeret2020-tmr}, based on linearized equations
that are valid for weak fields $h\to0$, and with an analysis of the
resulting magnetoelectric currents on length scales long compared to the coherence length $\xi_0$.
The latter approximation results to an oversimplification of the currents
when either $W$ or $L$ is of the order of $\xi_0$, discussed in more detail below.

We can now solve the self-consistent problem for the discretized
problem by minimizing the free energy in $Q$ and $\Delta$. We ignore
electromagnetic self-field effects, so the result is valid in the limit of negligible
magnetic screening, i.e., long Pearl length~\cite{pearl1964-cds}
$\Lambda=m/(e^2\mu_0n_s)\gg{}L,W$.

To characterize the strength of the Rashba spin-orbit interaction and
the magnetoelectric conversion, it is useful to introduce the following rates:
\begin{align}
  \Gamma_{r} &= 4 D \alpha^2
  \,,
  &
  \Gamma_{st} &= \Gamma_{r} \frac{\ell^2\alpha}{k_F\ell\xi_0} = \frac{\Gamma_r^{3/2}\Delta_0^{1/2}}{2E_F}
  \,,
\end{align}
where $\xi_0=\sqrt{D/\Delta_0}$ is the zero-temperature
coherence length.  Here, $\Gamma_{r}$ is the prefactor of the
Dyakonov--Perel spin-orbit relaxation term~\cite{efetov1980-idm}
$F=\ldots+\frac{i\pi\nu_FT}{8}\Tr[\frac{\Gamma_{r}}{2}\sum_{i=x,y}Q
  \sigma_i Q \sigma_i]$ that appears from the gradient term of the
action. Moreover,
$\Gamma_{st}=\frac{D\ell^2}{k_F\ell}4\alpha^3\xi_0^{-1}$ indicates the
strength of the singlet-triplet conversion cross-terms
$\frac{d\sigma_{xy}}{dB}F_{ij}[-i\alpha\sigma,Q]\nabla_jQ\propto{}\Gamma_{st}$.

\begin{figure}
  \includegraphics[width=\columnwidth]{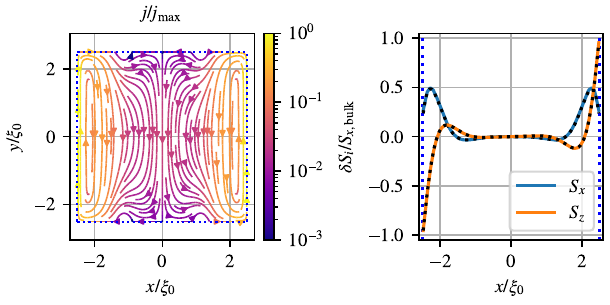}
  \caption{\label{fig:2dsccurrent-sq}
    Left: Current flow for
    $\Gamma_{r}=10\Delta_0$, $\Gamma_{st}=0.4\Delta_0$, $T=0.2\Delta_0$, $h=0.1\Delta_0$,
    $L=5\xi_0$, $W=5\xi_0$, with grid size $60\times60$.
    Line color indicates current amplitude $|\vec{j}|$.
    Right: Corresponding spin density oscillation $\delta S(x,y)=S(x,y)-S_{\rm bulk}$ at $y=0$.
    Dotted black line is the $h\to0$, $W\gg{}L$ limit analytical solution~\cite{bergeret2020-tmr}.
  }
\end{figure}

\begin{figure}
  \includegraphics[width=0.75\columnwidth]{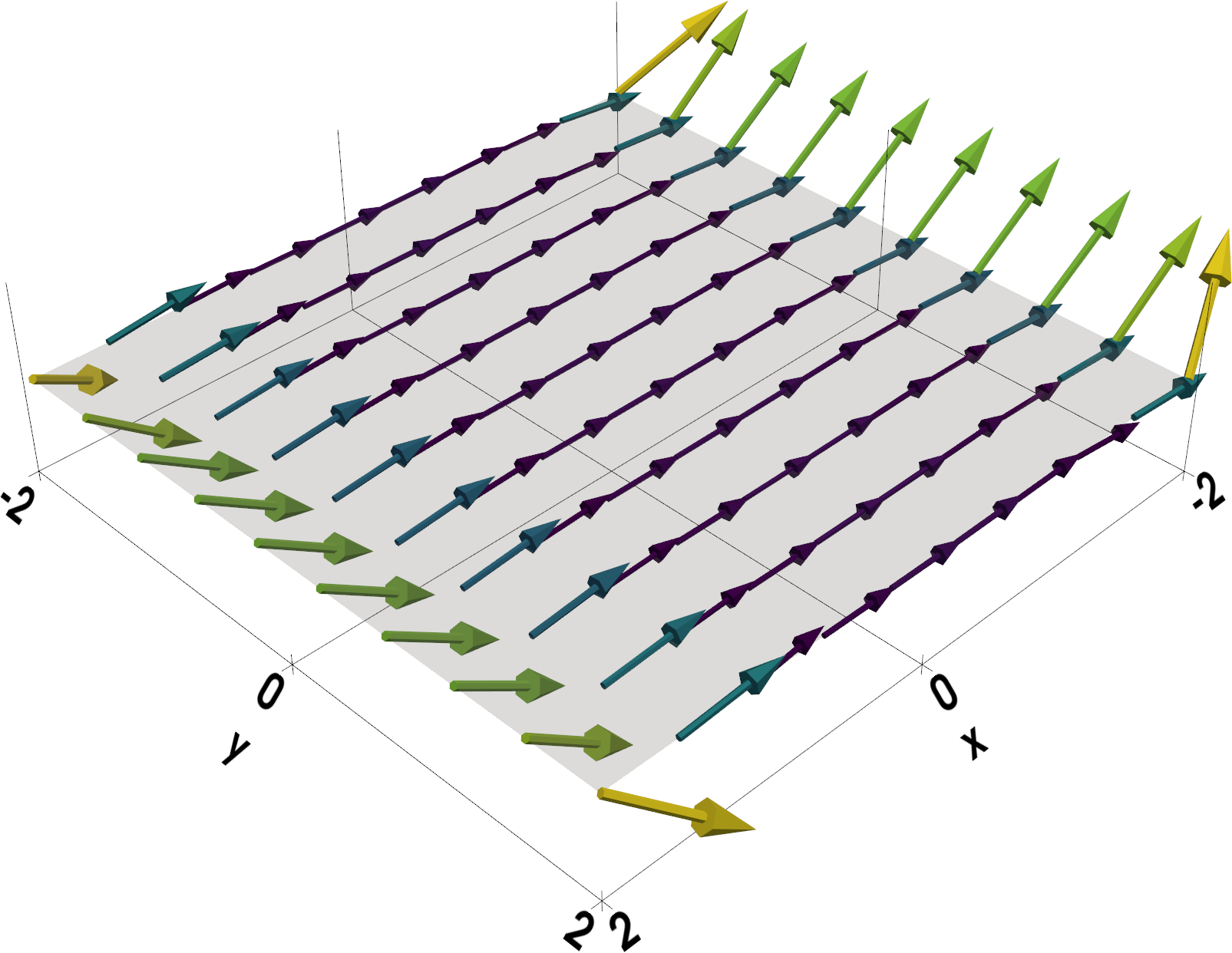}
  \caption{\label{fig:2dsS-sq}
    Spin texture corresponding to Fig.~\ref{fig:2dsccurrent-sq}.
  }
\end{figure}

\begin{figure}
  \includegraphics[width=\columnwidth]{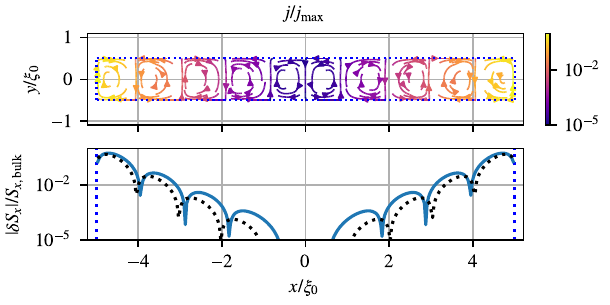}
  \caption{\label{fig:2dsccurrent}
    Same as Fig.~\ref{fig:2dsccurrent-sq}, but for $L=10\xi_0$, $W=\xi_0$,
    with grid size $220\times22$.
  }
\end{figure}

The calculated charge current and spin density $S_x$ for a square
sample $L=W=10\xi_0$ are illustrated in
Fig.~\ref{fig:2dsccurrent-sq}. The spin density follows the small-$h$
analytical result of Ref.~\onlinecite{bergeret2020-tmr} essentially
exactly, except in the corners where the spin texture tilts in $y$-direction,
as seen in Fig.~\ref{fig:2dsS-sq}.
The current is qualitatively similar to what is discussed in Ref.~\onlinecite{bergeret2020-tmr},
however note the appearance of four ``vortices'' close to $x=0$. This
structure becomes more apparent for $L\gg{}W\sim\xi_0$, as shown in
Fig.~\ref{fig:2dsccurrent}.

The result can be understood as follows.
The spin-dependent gauge field $A$ together with a sample boundary
generates a perturbation in the triplet component $\vec{Q}_t$ of
$Q=Q_s+\vec{Q}_t\cdot\vec{\sigma}$, and also in the spin density. The perturbation has an oscillatory decay
$\delta\vec{Q}_t\propto{}e^{-k_nx}$ from the boundary toward the bulk value, with the complex decay wave vectors~\cite{bergeret2020-tmr}
\begin{align}
  \label{eq:knpm}
  ik_{n}^\pm = \sqrt{\kappa_n^2 - 2\alpha^2 \pm 2i\sqrt{7\alpha^2 + 4\kappa_n^2}}
  \,,
\end{align}
where $\kappa_n^2=2\sqrt{\omega_n^2+\Delta^2}/D$. The $F_{ij}$ SOC term couples these (spin) oscillations to the singlet (charge) sector.
This is simplest to consider for weak superconductivity, $f,f^\dagger\to0$:
\begin{align}
  Q \simeq \begin{pmatrix} 1 - \frac{1}{2}f f^\dagger & f \\ f^\dagger & -1 + \frac{1}{2} f^\dagger f \end{pmatrix}
  \,.
\end{align}
Writing $f=f_s + \vec{f}_t\cdot\vec{\sigma}$, $f^\dagger=f_s^\dagger + \vec{f}_t^\dagger\cdot\vec{\sigma}$ and assuming Rashba interaction,
the singlet-triplet coupling term in the action becomes
\begin{align}
  F_{ST}
  &=
  \frac{\pi\sigma_{xy}'T}{4}\alpha^2
  \Tr[
    \partial_x f_t^z \partial_y f_s^\dagger - \partial_x f_t^{z\dagger} \partial_y f_s
    \\\notag&\qquad
    - 2\alpha(\vec{f}_t\times\nabla f_s^\dagger)\cdot\hat{z}  + 2\alpha(\vec{f}_t^{\dagger}\times\nabla f_s)\cdot\hat{z}
  ]
  \,.
\end{align}
In Ginzburg--Landau expansion, $f_{s/t}(\vec{r})=\Delta(\vec{r}) \tilde{f}_{s/t}$, $f_{s/t}^\dagger=\Delta(\vec{r})^* \tilde{f}_{s/t}$,
this produces a Lifshitz invariant~\cite{bauer2012-ncs}
\begin{align}
  F_{ST}
  &=
  i(\vec{\eta}\times{\hat{z}})
  \cdot
  (
  \Delta^* \nabla \Delta - \Delta \nabla\Delta^*
  )
  \,,
\end{align}
where $\vec{\eta}\propto\alpha^3\sum_{\omega_n}\tilde{\vec{f}}_t
\tilde{f}_s$. Then also $\vec{\eta}\approx\eta_y(x)\uvec{y}$
oscillates as a function of distance from the surface, producing an
effective magnetic field
$\vec{B}_{\rm{}eff}(x)=\uvec{z}\partial_x\eta_y(x)$. This then results
to the circulating currents visible in
Fig.~\ref{fig:2dsccurrent}. Increasing the strip width results to
averaging over these short-range oscillations, and gradually
transforms the solution towards uniform current flow as seen in
Fig.~\ref{fig:2dsccurrent-sq}.

\subsection{Inverse spin-galvanic effect}

The converse to the above is the inverse spin-galvanic effect (ISGE)
or Edelstein effect, where the singlet-triplet coupling generates a spin
density from a charge current.

To find ISGE in a uniform infinite 2D strip of width $W$, we can
assume a nonzero orbital field $A_x$ driving current along the strip
in $x$-direction. The Rashba spin-orbit coupling is as assumed in the previous section,
and we take $h=0$.  Then, $\hat{\nabla}_xQ=\partial_xQ - i[A_x\tau_3 +
  \alpha\sigma_y, Q]$, and we can assume $\Delta$ is real, and find
the solution $Q=Q(y)$ of the saddle-point equations. The saddle point
equations can be written as~\cite{virtanen2022-nsm}
\begin{align}
  \label{eq:usadel-leading}
  &D\hat{\nabla}_i(Q\hat{\nabla}_iQ)
  -
  [\tilde{\Omega}, Q]
  =
  -\tilde{\eta} \mathcal{T}
  \,,
  \\
  \mathcal{T}
  &=
  \hat{\nabla}_i J^H_i
  \,,
  \quad
  \tilde{\Omega}
  =
  \omega_n\tau_3 + \Delta\tau_1
  \,,
  \\
  J_i^H &=
  \{F_{ij}+QF_{ij}Q,\hat{\nabla}_jQ\} -i \hat{\nabla}_j(Q[\hat{\nabla}_iQ,\hat{\nabla}_jQ])
  \,,
\end{align}
where $\tilde{\eta}=\frac{D\ell^2}{4k_F\ell}=\frac{D\tau}{4m}=\frac{\xi_0\Gamma_{st}}{16\alpha^3}$.
Analytical results can be found by working in the leading order in
$\tilde{\eta}$ and $A_x$: assume $Q=Q_0 + \delta Q + \mathcal{O}(\tilde{\eta}^2,A_x^2)$, where
$Q_0=\tilde{\Omega}/\sqrt{\omega_n^2+\Delta^2}$ is the equilibrium bulk
solution, $\{Q_0,\delta Q\}=0$, and $\delta Q\propto{}A_x\tilde{\eta}$.
We have then $\hat{\nabla}_xQ_0=-iA_x[\tau_3,Q_0]=2A_xf_0\tau_2$ and $\hat{\nabla}_yQ_0=0$,
and $[F_{ij},Q_0]=0$. Also, $\delta Q$ does not appear in $\mathcal{T}$ in the leading order, so that
\begin{align}
  \mathcal{T}
  &\simeq
  2\{\hat{\nabla}_yF_{yx},\hat{\nabla}_xQ_0\}
  =
  -32\alpha^3A_xf_0\sigma_y\tau_2
  \,,
\end{align}
where $f_0=\Delta/\sqrt{\omega_n^2+\Delta^2}$.

For $W\to\infty$, we can assume $\partial_y\delta{}Q=0$, and $\delta{}Q\propto\sigma_y$.
The spin relaxation term in Eq.~\eqref{eq:usadel-leading} then obtains the form
$D\hat{\nabla}_i(Q\hat{\nabla}_iQ)\simeq-\Gamma_{r}Q_0\delta{}Q$.
The equation is solved by $\delta Q = [\Gamma_{r} Q_0 + 2\tilde{\Omega}]^{-1}\tilde{\eta}\mathcal{T}$, i.e.,
\begin{align}
  \label{eq:Qbulk}
  \delta Q_{\rm bulk}
  =
  \Gamma_{st} \xi_0 A_x \frac{
    i[\Delta\omega_n\tau_1 - \Delta^2\tau_3]\sigma_y
  }{
    (\frac{1}{2}\Gamma_{r} + \sqrt{\omega_n^2+\Delta^2})(\omega_n^2+\Delta^2)
  }
  \,.
\end{align}
The induced bulk spin density is $\vec{S}_{\rm bulk}=S_y\uvec{y}$, where
\begin{align}
  S_y = \Gamma_{st} \xi_0 A_x \pi\nu_F T\sum_{\omega_n}\frac{\Delta^2}{(\frac{1}{2}\Gamma_{r} + \sqrt{\omega_n^2+\Delta^2})(\omega_n^2+\Delta^2)}
  \,.
\end{align}
As shown in~\cite{edelstein1995-mep}, $\vec{S}$ is perpendicular both to
the Rashba direction $F_{ij}\propto\sigma_z$ and the current flow. 
This bulk value agrees with previous results~\cite{edelstein2005-med,konschelle2015-tsg,tokatly2017-uep}.

The vacuum boundary conditions at $y=\pm{}W/2$ read:
\begin{align}
  \label{eq:bccond}
  J_{y,\mathrm{tot}}
  =
  DQ(\partial_y Q - i[-\alpha\sigma_x, Q]) + \tilde{\eta} J_y^H = 0
  \,.
\end{align}
The bulk solution~\eqref{eq:Qbulk} does not satisfy it, which results
to a perturbation that decays towards the bulk similarly as seen in
the previous section. From the spin structure of the equations, one
can observe that the solution for $\delta Q$ in leading order in
$\tilde{\eta}$, $A_x$, should generally have the form
$\delta{}Q=\delta{}Q_{\rm{}bulk}+\tilde{Q}$,
$\tilde{Q}=\tilde{Q}_y\sigma_y+\tilde{Q}_z\sigma_z$.  Consequently,
spin accumulation $S_z\ne0$ can appear close to the strip edges,
whereas $S_x=0$.

We can proceed to solve the spin accumulation analytically.
Working again the leading order in $\tilde{\eta},A_x$,
the boundary condition and Eq.~\eqref{eq:usadel-leading} read
\begin{align}
  \label{eq:bceq}
  &D(\partial_y + 2i\alpha \sigma_x)\tilde{Q} = 16\tilde{\eta} A_x \alpha^2 f_0 Q_0 \sigma_z \tau_2 - 2iD\alpha\sigma_x\delta Q_{\rm bulk}
  \,,
  \\
  &D(\partial_y +  2i\alpha\sigma_x)^2\tilde{Q} - \Gamma_{r}\frac{\tilde{Q} - \sigma_y\tilde{Q}\sigma_y}{2}
  \\&\qquad\notag
  - 2\sqrt{\omega_n^2+\Delta^2} \tilde{Q} = 0
  \,.
\end{align}
The second equation has the general solution $\tilde{Q}=\sum_{ab=\pm}\tilde{Q}_{zab}[\sigma_z + B_n^b \sigma_y] e^{iak_n^b y}$,
$B_n^b = -4\alpha ik_n^b / [(ik_n^b)^2 - 4\alpha^2 - \kappa_n^2]$, where $k_n^\pm$ are given by Eq.~\eqref{eq:knpm}.
The coefficients $\tilde{Q}_{zab}$ are determined by Eq.~\eqref{eq:bceq} at $y=\pm{}W/2$, and can be solved.

\begin{figure}
  \includegraphics[width=\columnwidth]{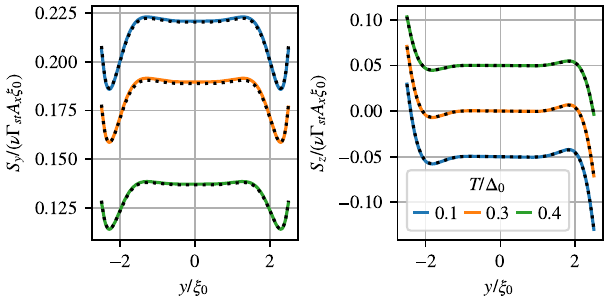}
  \caption{\label{fig:isge}
    Inverse spin-galvanic effect. Spin density at different temperatures
    is shown, for $A_x\xi_0=0.005$, $W=5\xi_0$, $L=4\xi_0$, $\Gamma_{r}=10\Delta_0$, and $\Gamma_{st}=0.4\Delta_0$,
    with grid size $90\times90$.
    Left: $S_y$. Right: $S_z$ (curves offset vertically by $\pm0.05$ for clarity).
    Solid lines indicate numerical results, black dotted lines solutions to
    Eq.~\eqref{eq:bceq}.
  }
\end{figure}

The resulting spin density and corresponding numerical results
are shown in Fig.~\ref{fig:isge}, for a finite-size case where
$L,W>\xi_0$. At the left and right ends, we assume $Q$ is fixed to
the values $Q_0\rvert_{\Delta\mapsto{}\Delta e^{\pm{}iA_xL/2}}$ which
in the self-consistent calculation generates a nearly uniform phase
gradient along $x$. The numerical and analytical results for the spin
density generated by the inverse galvanic effect agree for the small phase gradient
$A_x$ chosen.

Spin accumulation $S_z\ne0$ at the boundaries in general could be expected
to be generated by the spin-Hall effect~\cite{sinova2014-she}. However, in the leading order
in $\tilde{\eta}$, the total matrix current in the bulk is
$J_{y,\mathrm{tot}}\propto\tau_2\sigma_z$, and from
Eq.~\eqref{eq:jspin} one finds $J^s_{ij}=0$, i.e., no bulk spin
current. The behavior here is then somewhat different from a spin-Hall effect,
where bulk spin current is balanced by boundary spin accumulation and
relaxation. In addition, in the system here $S_z\ne0$ only in the superconducting state.
In dirty 2D Rashba metals with a parabolic spectrum
in the normal state the DC spin-Hall effect vanishes,
and $S_z=0$ at strip boundaries.~\cite{inoue2004-sps,mishchenko2004-scp}
This also follows from the
present theory in the normal state: under similar approximation as above
using the normal-state nonequilibrium form of the equations,~\cite{virtanen2022-nsm,tokatly2017-uep} one finds
$J_{y,\mathrm{tot}}=0$ which also implies the vanishing spin-Hall
effect~\cite{hijano2023-dhr}.

\section{Summary and conclusions}\label{sec:summary}

We presented an approach where numerical saddle-point solutions to the
spin-orbit coupled diffusion theory in 2D superconductors are obtained
relatively directly from the original action formulation. The
formulation is constructed in such a way that it satisfies exact
discrete versions of the conservation laws originating from symmetries
in the original theory. We applied it to studying the spin-galvanic
effects in finite-size Rashba superconductors, where the SOC and
exchange field or charge currents interact with the sample boundaries,
generating oscillations in the spin density. We considered these
problems also analytically, and find that the numerical and analytical
results fully agree in the validity range of the latter.  For the
inverse spin-galvanic effect, we find that equilibrium supercurrent
does generate boundary spin accumulations qualitatively similar to the
spin-Hall effect, even though quasiparticle current in the same system
in the normal state does not.

The approach is not limited to the specific model and effects considered above.
It can be applied also to studying the superconducting diode
effect in the dirty limit~\cite{ilic2024-sde}, or spin-orbit effects in multiterminal structures. A
nonequilibrium version can be used to model the quasiparticle
spin-Hall effect~\cite{sinova2014-she}, and its interaction with other
nonequilibrium effects in spin-split superconductors.~\cite{bergeret2018-cne} Moreover,
Refs.~\onlinecite{schwiete2021-nsm,virtanen2022-nsm} discuss extensions of
the theory to include thermoelectric effects from electron-hole
asymmetry, the implications of which in this framework are so far
not studied in superconductors. Ref.~\onlinecite{virtanen2022-nsm} also derives other
higher-order electron-hole symmetry breaking terms that are allowed
and in principle present already for the parabolic electron
dispersion, and may be important when the leading terms vanish. More
generally, one can also study effective actions where all symmetry-allowed
terms are included.~\cite{kokkeler2024-upc} The approach here can be
straightforwardly adapted to solving such actions numerically on the
saddle-point level, without requiring much manual work.

The computer program used in this manuscript is available~\cite{code}.

\acknowledgments%

I thank S. Ili\'{c}, F. S. Bergeret, and T. T. Heikkil\"a for discussions.
This work was supported by European Union's HORIZON-RIA program (Grant Agreement No.~101135240 JOGATE).

\bibliography{refs}

\appendix

\section{Matsubara formulation}\label{sec:matsubara}

Equation~\eqref{eq:action} is written in a Matsubara formulation
suitable for equilibrium problems. One way to obtain it from the
nonequilibrium Keldysh results in Ref.~\onlinecite{virtanen2022-nsm},
is to make an analytic continuation in the saddle point equations, and then find a free energy that generates
them. By construction the result has a similar form as the Keldysh action.
Another way is to re-do the derivation of the free energy from the
beginning in a Matsubara formulation.

For completeness and to define notation, let us outline the main
points in the second, replica $\sigma$-model approach.~\cite{finkelshtein1983-ici,belitz1994-amt} The problem is to evaluate
the disorder-averaged free energy $F=-T\langle\langle \ln Z
\rangle\rangle$, $Z=\tr{}e^{-H/T}$, corresponding to the
single-particle BdG Hamiltonian $\mathcal{H}$.  Random gaussian distributed disorder is assumed, and the
impurity field is correlated as $\langle\langle V_{\rm imp}(\vec{r})
V_{\rm{}imp}(\vec{r}')\rangle\rangle=\frac{1}{2\pi\tau_0\nu_F}\delta_{\vec{r}\vec{r}'}$,
where $\tau_0=\ell/v_F$ is the impurity scattering time.

The free energy is found via replica limit
$F=-T\lim_{N\to0}[\langle\langle{}Z^N\rangle\rangle-1]/N$.  The
disorder average of the partition function replicated $N$ times is
transformed to an integral over an auxiliary matrix field $Q$:~\cite{efetov1980-idm,finkelshtein1983-ici,belitz1994-amt}
\begin{align}
  \langle\langle Z^N \rangle\rangle
  &=
  \int\DD{\bar{\psi},\psi,V_{\rm imp}}
  e^{S[\bar{\psi},\psi]}
  e^{-\pi\tau_0\nu_F\int\dd{r} V_{\rm imp}(\vec{r})^2}
  \notag
  \\
  &\simeq
  \int\DD{\bar{\psi},\psi,Q} e^{S[Q,\bar{\psi},\psi]}
  \,,
  \\\notag
  S[\bar{\psi},\psi]
  &=
  -\int\dd{r}\int_0^{1/T}\dd{\tau}\sum_{\alpha=1}^N
  (\bar{\Psi}^\alpha_{\vec{r}\tau})^T\tau_3[\partial_\tau + \mathcal{H}] \Psi^{\alpha}_{\vec{r}\tau}
  \,,
  \\\notag
  \label{eq:S2}
  S[Q,\bar{\psi},\psi]
  &=
  \int\dd{r}
  \bar{\Psi}_{\vec{r}}^T[i\omega\tau_3 - \tau_3\check{\mathcal{H}}_0 + \frac{i}{2\tau_0}Q]\Psi_{\vec{r}}
  \\&\qquad
  - \frac{\pi\nu_F}{8\tau_0}\int\dd{r} \tr Q^2
  \,.
\end{align}
Here, the Nambu--spin column vectors are
$\Psi_{\vec{r}\tau}^\alpha=(\psi_{\uparrow\vec{r}\tau}^\alpha,\psi_{\downarrow\vec{r}\tau}^\alpha,\bar{\psi}_{\downarrow\vec{r}\tau}^\alpha,-\bar{\psi}_{\uparrow\vec{r}\tau}^\alpha)/\sqrt{2}$
and
$\bar{\Psi}_{\vec{r}\tau}^\alpha=(\bar{\psi}_{\uparrow\vec{r}\tau}^\alpha,\bar{\psi}_{\downarrow\vec{r}\tau}^\alpha,-\psi_{\downarrow\vec{r}\tau}^\alpha,\psi_{\uparrow\vec{r}\tau}^\alpha)/\sqrt{2}$,
and contain the electron Grassmann fields for each of the replicas
$\alpha=1,\ldots,N$ and spins $\uparrow,\downarrow$.
In Eq.~\eqref{eq:S2} they are represented as vectors of imaginary-time Fourier components
$\Psi_{\vec{r}m}=\sqrt{T}\int_0^{1/T}\dd{\tau} e^{i\omega_m\tau}\Psi_{\vec{r}\tau}$,
$\bar{\Psi}_{\vec{r}m}=\sqrt{T}\int_0^{1/T}\dd{\tau}e^{-i\omega_m\tau}\bar{\Psi}_{\vec{r}\tau}$, $\omega_m=2\pi T(m+\frac{1}{2})$.
They are related by $\bar{\Psi}_{\vec{r}}=C\Psi_{\vec{r}}$, where the charge conjugation matrix is $C_{mm'}^{\alpha\alpha'}=-i\sigma_y\tau_1\delta_{\alpha\alpha'}\delta_{\omega_m,-\omega_{m'}}$. The matrix-vector products and $\tr$ imply summations over $m$,$\alpha$
and the Nambu--spin structure.
Note that to make connection with quasiclassical Green function theory, we use here a different
convention than Refs.~\onlinecite{finkelshtein1983-ici,belitz1994-amt}, e.g. in the lower Nambu block
$\omega_m\mapsto-\omega_m$. This results to some differences in factors of $\tau_3$ and
in the definition of $C$.
In Eq.~\eqref{eq:S2}, the matrix $(\check{\mathcal{H}}_0)^{\alpha\alpha'}_{mm'}=\delta_{\alpha\alpha'}T\int_0^{1/T}\dd{\tau}e^{i(\omega_m-\omega_{m'})\tau}\mathcal{H}_0(\tau)$
contains the BdG Hamiltonian from Eq.~\eqref{eq:Hstart}.
Finally, the integral over the field $Q=Q^{\alpha\alpha'}_{mm'}(\vec{r})$ is defined with the constraints $Q(\vec{r}) = Q(\vec{r})^\dagger = C Q(\vec{r})^T C^T$.

Integration over $\bar{\psi}$, $\psi$ gives
\begin{align}
  \label{eq:SQ}
  S[Q]
  &=
  \frac{1}{2} \Tr \ln[i\omega\tau_3 - \tau_3\check{\mathcal{H}}_0 + \frac{i}{2\tau_0}Q]
  -
  \frac{\pi\nu_F}{8\tau_0}\Tr Q^2
  \,,
\end{align}
where $\Tr$ includes also the integration over $\vec{r}$.  This action
at $\check{A}=0$, $\Delta=0$ is known to have spatially uniform saddle
point configurations $Q=V\Lambda{}V^\dagger$ where unitary $V$ has
symmetry compatible with $Q$.  The replica-symmetric saddle point is
$\Lambda_{mm'}^{\alpha\alpha'}=\delta_{\alpha\alpha'}\delta_{mm'}\sgn(\omega_m)\tau_3$.
Diffusion theory is found by making a gradient expansion by
considering $V_{\vec{r}}$ varying slowly in space, and expanding in
$1/\mu$ and the mean-free path $\ell$.~\cite{efetov1980-idm,belitz1994-amt} Expansion with
spin-orbit fields $\check{A}$ in $\mathcal{H}_0$ was done in
Ref.~\onlinecite{virtanen2022-nsm}.  The procedure there makes no
reference to the matrix structure of $Q$, only requiring that
$\Lambda^2=1$ and that
$\tau_3\check{\mathcal{H}}_0\rvert_{\Delta=0,\check{A}=0}$ is
proportional to the identity matrix and has the free-electron
form. Both hold here, and so the same algebraic steps apply and give
formally the same result.

Hence, the gradient expanded action becomes:
\begin{align}
  \label{eq:SQgradexp}
  S[Q]
  &\simeq
  -\frac{\pi}{8}\Tr[\frac{\sigma_{xx}}{2}(\hat{\nabla}Q)^2 + 4i\nu_F\Omega Q
    \\\notag&\qquad
    - \frac{\sigma_{xy}'}{2} F_{ij}Q\hat{\nabla}_iQ\hat{\nabla}_jQ]
  \,,
\end{align}
with the constraint $Q_{\vec{r}}^2=1$. It differs from the Keldysh
result only in the matrix structure, in agreement with the analytic
continuation argument.

When going from Eq.~\eqref{eq:SQ} to Eq.~\eqref{eq:SQgradexp}, parts
of the action that do not depend on gradients of $V$ are
discarded. This ``high-energy'' part
(cf.\ e.g.~\cite{kamenev2010-kta}) contains terms producing
the equilibrium response of normal-state electron gas.
It includes the term
\begin{align}
  S_h
  &=
  -\frac{1}{16}\Tr (\mathcal{G}\mathcal{A}_0)^2
  \simeq
  \frac{N\nu_F}{4T}\int\dd{r}(\mathcal{A}_0)^2
  =
  -\frac{N F_h}{T}
  \,,
\end{align}
where
$\mathcal{G}^{-1}=i\omega\tau_3+\mu-\vec{p}^2/(2m)+\frac{i}{2\tau_0}\Lambda$,
and $k_F\ell\gg1$.  From this, one finds
$S_P^a=-\frac{\delta}{\delta\mathcal{A}_0^a}F_h=\frac{1}{2}\nu_F\mathcal{A}_0^a$,
the Pauli paramagnetic contribution, so that the total spin density is
the sum of $S_P^a$ and the spin accumulation~\eqref{eq:Sdef}. A
similar conclusion holds for the normal-state equilibrium
currents~\cite{tokatly2008-esc}.

\subsection{Saddle point}

The classical saddle point is scalar in replicas,
$Q_{\vec{r}mm'}^{\alpha\alpha'}=\delta_{\alpha\alpha'}Q_{\vec{r}mm'}'$.
In this case, $S[Q] = N S'[Q']$ where $S'$ is the action of a single
replica. In the saddle-point approximation for the integration over
$Q$, the free energy is then
\begin{align}
  F_0 &\simeq{} -T S'[Q']
  \,.
\end{align}
In the main text, we work in this approximation only, and drop the
replica structure.  When the fields $\check{A}$, $\Delta$ do not
depend on imaginary time and noninteracting problem is considered as
above, $Q'$ is also diagonal in the Matsubara frequencies. This then
leads to Eq.~\eqref{eq:action}.

The symmetries of the saddle-point value reflect symmetries of the
action. Equation~\eqref{eq:SQ} has symmetries $S[Q]=S[CQ^TC^T]$, and
$S[Q]^*=S[-\tau_3 u Q^\dagger u \tau_3]$ where
$u_{mm'}=\delta_{\omega_m,-\omega_{m'}}$.  With $Z$ being real-valued,
these result to $Q=CQ^TC^T$ and $Q=-\tau_3uQ^\dagger u \tau_3$ at the
saddle point. The constraint $Q=Q^\dagger$ does not need to hold, as
the integration contours are deformed in the
complex plane to pass through the saddle point.  For
Matsubara-diagonal $Q$, these symmetries imply
Eqs.~\eqref{eq:Qsymmetry}.

\end{document}